\documentclass{article}

\usepackage{arxiv}

\usepackage[utf8]{inputenc} 
\usepackage[T1]{fontenc}    
\usepackage{hyperref}       
\usepackage{url}            
\usepackage{booktabs}       
\usepackage{amsfonts}       
\usepackage{nicefrac}       
\usepackage{microtype}      
\usepackage{lipsum}		
\usepackage{graphicx}
\usepackage{natbib}
\usepackage{doi}
\usepackage{subcaption}

\title{Implications of the Pessimistic Lower Limit on the Drake Equation}

\author{ \href{https://orcid.org/0009-0004-0055-5476}{\includegraphics[scale=0.06]{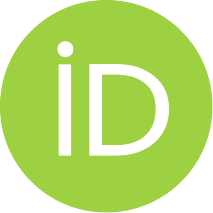}\hspace{1mm}Max Baak} \\
	Unaffiliated\\
	Amsterdam \\
	The Netherlands \\
	\texttt{max.baak@gmail.com} \\
	\And
	\href{https://orcid.org/0000-0003-4579-2120}{\includegraphics[scale=0.06]{orcid.pdf}\hspace{1mm}Hella Snoek} \\
	University of Amsterdam\\
    Nikhef\\
	Amsterdam\\
	The Netherlands \\
	\texttt{h.l.snoek@uva.nl} \\
}

\renewcommand{\shorttitle}{Implications of the Pessimistic Lower Limit on the Drake Equation}

\hypersetup{
pdftitle={Implications of the Pessimistic Lower Limit on the Drake Equation},
pdfsubject={physics.pop-ph},
pdfauthor={Max Baak, Hella Snoek},
pdfkeywords={Drake equation, abiogenesis, old evidence problem, counting experiment, statistical inference, Fermi paradox, SETI},
}

\begin{document}
\maketitle

\begin{abstract}

The observation of life on Earth is generally accepted to be uninformative concerning the probability of life on other Earth-like planets, a belief first formalized by Brandon Carter and based on the selection effect of our existence.
In a similar way, the Drake equation is either presented as an estimate of the total number of active, communicative, extraterrestrial civilizations in our Galaxy ($n^g_{\rm civ}$), i.e. excluding humanity, 
or humanity is included in the estimate but judged to be an uninformative data point.
The literature speaks of pessimism lines: if $n^g_{\rm civ}$ is below a certain value, humanity is unique in the Galaxy (or observable universe), 
however no statistical lower limit on $n^g_{\rm civ}$ is defined based on humanity's existence.

Daniel Whitmire has recently challenged the Carter abiogenesis argument, 
claiming the logic behind it is flawed, as
the conditional likelihoods used by Carter in Bayes' theorem are not evaluated prior to the occurrence of the evidence of life on Earth, but posterior.
Doing so correctly, 
the anthropic selection effect is removed and the observation of life on Earth is informative about 
the probability of life after all.

Following this argument, 
we treat the Drake equation as estimate of all technological civilizations in a statistical counting experiment and include the data point of humanity as informative evidence.
This allows one to set a pessimistic lower limit on $n^o_{\rm civ}$ for the observable universe, $n^o_{\rm civ} > 0.051$ at 95\% C.L., or $n^g_{\rm civ} > 8\times10^{-13}$ at 95\% C.L. for the Galaxy.
In particular,
this excludes models that predict $n^o_{\rm civ}\ll 1$ for the observable universe
and refines the allowable parameter space for hypotheses like Rare Earth.
Our analysis substantially reduces the portion of the Drake equation parameter space that predicts humanity is alone;
when applying the lower limit
this study finds $P(n^o_{\rm civ}>1 |\, {\rm humanity}) = 97.6\%$, 
making solitude in the observable universe a disfavored outcome.
For the low-end estimate of $n^o_{\rm civ}\! =\! 1$ we calculate a probability of 42\% for the existence of other communicating civilizations.

\end{abstract}


\section{Introduction}

In two recent papers~\cite{whitmire2023abiogenesis, whitmire2025abiogenesis} refutes the Carter abiogenesis argument, as summarized here.

The Carter abiogenesis argument~(\cite{carter1983anthropic}) is based on the selection effect implied by our existence: humanity finds itself on a planet where abiogenesis must have taken place, and therefore, it is argued, this observation alone provides no information about the actual probability of abiogenesis.
This argument has been widely accepted in the literature~(\cite{crick1981life, bostrom2013anthropic, spiegel2012bayesian, mash1993big, kukla2010extraterrestrials, waltham2014lucky, walker2017origins}). 

According to Carter, life on Earth (LoE) exists regardless of the hypotheses whether abiogenesis is easy (AB easy) or hard (AB hard), i.e. one has the conditional likelihoods $P({\rm LoE}|{\rm AB\ easy}) = P({\rm LoE}|{\rm AB\ hard}) = 1$. 
Consequently, filling in Bayes' theorem and using $P({\rm AB\ easy}) + P({\rm AB\ hard}) = 1$, 
one has: $P({\rm AB\ easy}|{\rm LoE}) = P({\rm AB\ easy})$. 
In other words, the evidence of life on Earth is not informative as it does not update the prior.
One could call this the fate principle, in the sense that our existence is uninformative if one assumes it has always been a given.
Note that, if correct,  Carter implies that once evidence exists, the conditional likelihoods are equal to 1, independent of the hypotheses, and the evidence is thus predictively irrelevant.

Whitmire challenges the Carter argument by drawing upon the ``old evidence problem'' from Bayesian confirmation theory~(\cite{morton1981clark}), and claims 
the evidence of life on Earth is mistakenly judged as uninformative.
%
%
To evaluate a model, it does not matter whether or not the evidence has already manifested itself. 
In the case of old evidence, i.e. evidence known previously, such as life on Earth, 
the conditional likelihoods in Bayes' theorem should be evaluated counterfactually or ``historically'': not posterior to the occurrence of the evidence, as done by Carter, but prior to the occurrence.

The old-evidence argument is readily illustrated with a coin-toss example, with the existing evidence that a coin has landed heads $N$ times in a row. 
To infer if the coin is fair or unfair,
the correct analysis should not use the (uninformative) posterior conditional probabilities of 1,
but the ones prior to the coin flip:
$(1/2)^N$ for a fair coin and $1$ for an unfair one with two heads.
See~\cite{whitmire2023abiogenesis} for more good examples.

Taking the historical evaluation into account, the observation of life on Earth is informative and the anthropic selection effect is removed: before abiogenesis occurred, Earth was indistinguishable from any other prebiotic Earth-like planet and can be viewed as a representative example of the broader class of all Earth-like planets.
Furthermore, any conclusions about abiogenesis on Earth should also apply to this reference class; 
notwithstanding the fact that humanity finds itself on Earth where abiogenesis has necessarily taken place.

In this work the old-evidence argument is employed to set a statistical lower limit on the Drake equation and to evaluate the corresponding implications.
In doing so, we do not assume or argue that life is widespread, e.g. unlike the Principle of Mediocrity. 
The number of Earth-like planets in the Galaxy or observable universe may be limited, and the probability of abiogenesis may be tiny.
Although the observation of life on Earth is informative, it is a single data point, and thus any inference derived from this instance has limited significance.
Regardless, the corresponding lower limit 
provides useful information for taking a position on the question of extraterrestrial life in the observable universe.

We define $n_{\rm civ}^{g}$ and $n_{\rm civ}^{o}$ as the estimates of the Drake equation for the Galaxy and observable universe respectively.
%
%
With the view that the evidence of humankind is uninformative or
the Drake equation applies only to extraterrestrial civilizations,
models with $n^g_{\rm civ}\ll1$ ($n^o_{\rm civ}\ll1$)
suggest we are likely alone in the Galaxy (observable universe). 
The literature defines pessimism lines, where if $n^g_{\rm civ}$ ($n^o_{\rm civ}$) is below a certain value, humanity is unique in the Galaxy (observable universe), and if above, other technological civilizations must exist or have existed.
Using the inclusive interpretation advocated here, 
model predictions need to accommodate the observed number of communicating civilizations, 
%
i.e. $n_{\rm obs}\! \ge\! 1$. 
Estimates of $n^o_{\rm civ}\!\ll\!1$ and $n^g_{\rm civ}\!\ll\!1$ are therefore statistically constrained 
-- they may not rule out the existence of humanity --
and existing model predictions may need to be updated accordingly.

In addition, the lower limit on $n^o_{\rm civ}$ increases the probability of the existence of other communicating civilizations. 
%
The probability of any such occurrence is governed by counting statistics.
The chance of one event when expecting close to none is close to zero.
However, the odds of having a second event are much higher given the occurrence of a first one and a necessarily compatible expectation value -- 
for example, when using the baseline estimate of 1.



Our contributions are straightforward and twofold.
%
%
a) We define a pessimistic lower limit for the existence of communicative civilizations in the observable universe.
%
b) We demonstrate the statistical implications thereof, namely
a restriction of the parameter space of models with $n^o_{\rm civ}\ll 1$, resulting in favorable odds for the existence of other technological civilizations in the observable universe.


\section{Related Work}

Throughout the literature, the Drake equation~(\cite{drake1965radio}) and the data point of humankind are not described very consistently. 
The Drake equation is often discussed within the contexts of searches for extraterrestrial intelligence (SETI)
or the Fermi paradox,
%
and presented there
as a function to estimate the number of extraterrestrial civilizations in our Galaxy that can communicate with Earth~(\cite{valamontes2025critical, prantzos2013joint, maccone2010statistical}).
Also done regularly, the Drake equation is correctly presented as including humankind; however, the evidence of humankind is treated as uninformative, either implicitly or explicitly, regarding the equation's low-end predictions~(\cite{molina2019searching, sandberg2018dissolving}).

\citet{frank2016new} ask the question: has even one other technological species ever existed in the observable universe, and set a limit on the probability that other technological species have ever evolved. 
Based on humanity's existence, they argue a pessimism line of $f_{bt} \sim 2.5\times 10^{-24}$ for 
the probability that a habitable zone planet develops a technological species.
If $f_{bt}$ is smaller than this value, then humanity is the only technological civilization that has evolved in the observable universe. 
If $f_{bt}$ is larger, then other technological civilizations should have evolved as well.
However, the authors do not present this as a statistical lower limit on $f_{bt}$, nor do they argue that the value applies to humanity as well.
In this paper we expand upon this work.

\citet{balbi2023beyond} perform a Bayesian analysis aiming to set a lower limit on the probability of abiogenesis on Earth-like planets ($p_{LE}$), based on the existence of humanity and depending on the (unknown) number of such planets in the Galaxy ($N_E$). 
For the lower limit on $p_{LE}$
they find a strong dependency on the value for $N_{E}$ and the prior distribution for $p_{LE}$.
Essentially there is insufficient information to constrain $p_{LE}$ alone,  
however no lower limit is set on the expected number of occurrences, $p_{LE}N_E$, which is constrained in this work.

\citet{sandberg2018dissolving}
include uncertainties on all factors in Eq.~\ref{eq:drake1} and sample the resulting values for $n^g_{\rm civ}$,
to find values many orders of magnitude lower than the inverse of the total number of stars 
in the observable universe.
Based on their study they derive 52\% probability for being alone in the Galaxy, and 38\% probability for being alone in the observable
universe.
These numbers increase (in a model-dependent way) when including the Fermi observation that no extraterrestrial civilizations have yet been observed. 
However, the authors ignore the information of $n_{\rm obs} \ge 1$, which constrains $n^g_{\rm civ}$ from below. 
%
Including this, as done here, considerably weakens the conclusion that we are likely alone in the observable universe.

We are not aware of similar research that 
places a pessimistic lower limit on $n^o_{\rm civ}$ or $n^g_{\rm civ}$ and evaluates the corresponding implications.


\section{The Drake Equation}

The Drake equation assumes that the evolution of technological civilizations is associated with planets and their host stars,
and is usually presented in the following form:
\begin{equation} \label{eq:drake1}
    n^g_{\rm civ} =  R^{*}\,f_p\,n_e\,f_l\,f_i\,f_c\,L\,,
\end{equation}
where the terms have the following meaning:
\begin{enumerate}
    \item[$R^*$]: rate of formation of stars in the Galaxy,
    \item[$f_p$]: fraction of those stars with planets,
    \item[$n_e$]: average number of planets per star in the habitable zone, with an environment suitable to life,
    \item[$f_l$]: fraction of planets on which life actually emerges,
    \item[$f_i$]: fraction of those planets on which intelligent life appears,
    \item[$f_c$]: fraction of civilizations that develop communication technology emitting detectable signals of existence into space, thus revealing their existence,
    \item[$L$]: length of time these civilizations emit such detectable signals.
\end{enumerate}

An alternative formulation scales instead with the number stars in the Galaxy, $N^*$,
and replaces $L$ with a fraction of existence $f_L$. 
A popular example is the Rare-Earth hypothesis~(\cite{wardrare}), its premise being that Earth-like planets are rare.
%
The Rare-Earth estimate includes five extra fractions that narrow down the habitable planets to Earth-like ones.
The hypothesis mostly predicts small values for $n^g_{\rm civ}$, in particular $n^g_{\rm civ}\ll1$.

Many other variants of the Drake equation exist~(\cite{westby2020astrobiological, hetesi2006new, maccone2012statistical, maccone2015statistical}). 
These versions typically add (or change) fractions to the formula that refine the conditions under which life may emerge. 

In practice, the parameters in the Drake equation should be considered averages, 
for example, encapsulating possible spatio-temporal conditions. 
One can also conceive the Drake equation to be a sum of predictions for Earth-like and other types of habitable planets, each class with different parameter values.

The Drake equation has no clear estimates. 
Many of its terms, such as $f_l, f_i, f_c, L$, are highly uncertain, with logarithmic ranges that can span tens of orders of magnitude. 
Likewise, predictions of $n^g_{\rm civ}$ range over many orders of magnitude, from non-existent to present for almost every star~(\cite{sandberg2018dissolving}). 
Basically, there are two prediction schools: $n^g_{\rm civ}\!\sim\! R^{*} L$ and $n^g_{\rm civ}\!\le\!1$. 
Notably, some estimates of $n^g_{\rm civ}$ are many orders of magnitude lower than the inverse of the total number of stars 
in the observable universe~(\cite{koonin2014origins, totani2020emergence, scharf2016quantifying}). 

Let us emphasize that the Drake equation predicts the total number of active, communicative
civilizations in the Galaxy, not just the number of extraterrestrial
ones.
The Drake equation, or one of its variants, lists the factors affecting the likelihood that a technological civilization makes its presence known on a habitable planet through 
emitting detectable signals.
In essence it describes a fraction of the Galaxy; 
starting from all born (or available) stars, it narrows down the selection factor by factor.
The remaining civilizations are all active, communicative and detectable.

Every factor of the Drake equation 
accommodates for Earth and humankind; 
none exclude them.\footnote{Even under the (strict) definition that technological civilizations must be detectable on Earth, by us, still that does not exclude humanity.}
The notion that the formula somehow does not apply to humankind
goes against its design principle that intelligent life can develop on any habitable planet given the right (albeit unknown) conditions.

From a statistics perspective, 
the search for communicating civilizations is
a counting experiment, comparing the estimated and observed numbers.
With life on Earth included in the Drake estimate, 
the instance of humanity must also be included in the observed total count, in line with Whitmire's argument.

In the formulation adopted here, based on $N^{*}$,
all factors to the right of $n_e$ are grouped into a single probability parameter, $p_L$, scaling with the total number of planets in the habitable zone, $N_H^g$:
\begin{eqnarray}\label{eq:drake3}
    n^g_{\rm civ} &=& N^{*}\,f_{p}\,n_{e}\,p_{L} \nonumber \\
    &=& N_H^g\,p_{L}\,.
\end{eqnarray}
Recognizing there are many more stars beyond our Galaxy, a notable alternative, relevant here, predicts the total number of communicating civilizations in the observable universe.
This dramatically increases $N^*$, from $3\times 10^{11}$ to $2\times 10^{22}$ stars~(\cite{silburt2015statistical}).\footnote{The James Webb space telescope has found a higher-than-predicted abundance of luminous galaxies in the early universe, suggesting refinements required to the
current galaxy formation models (an area of ongoing research), but not a breakdown of these theories or their predictions.} 
The number of habitable planets in the observable universe, $N^o_H$, 
is assumed to scale likewise, and,
assuming the same value for $p_L$, so does $n^o_{\rm civ}$.

\section{Pessimistic Lower Limit}

\citet{frank2016new} derive the pessimism line where Earth is the only location in the history of the cosmos where a technological civilization has ever evolved, namely $f_{bt} = f_l\,f_i\,f_c \sim 2.5\times 10^{-24}$.
They do not consider active technological civilizations; the term $L$ in Eq.~\ref{eq:drake1} is explicitly ignored.
The pessimism line is obtained from a simulation thought experiment. 
From a statistical perspective, with this value of $f_{bt}$, if one were to rerun the history of the Universe 100 times to its current state, only once would a  technological species ever have evolved on one of the habitable planets in the observable universe.
Pessimism lines are also set for the Galaxy and clusters of galaxies.

Life on Earth is implicitly taken as uninformative by the authors regarding the probability of abiogenesis on other worlds:
$f_{bt}$ is only presented as a pessimism line for the existence of other technological civilizations, and not as a statistical lower limit applicable to all civilizations, including humanity. 
(If the Drake equation were to apply to extraterrestrial civilizations only, 
values of $f_{bt} < 10^{-24}$ would be perfectly acceptable given the lack of evidence thereof.)

We extend this work in two ways.
First, $f_{bt} > 2.5\times 10^{-24}$ is a statistical lower limit at 99\% C.L. on all technological civilizations in the observable universe. 
This is clear from the simulation experiment: in 99 out of 100 reruns not a single technological civilization has ever evolved,
i.e. for smaller values of $f_{bt}$ the existence of humanity is statistically excluded.

We argue the lower limit on $f_{bt}$ should not be based on the Galaxy, a cluster of galaxies, or other limited group of stars.
The number of technological civilization that ever evolved scales with the number of suitable stars and habitable planets.
Clearly, using all available stars in the observable universe sets the most general, pessimistic lower limit.

Second, Frank and Sullivan's approach is extended to all \textit{active} technological civilizations in the observable universe. 
To do so, the simulation example is reused.
If one were to rerun the evolution of the Universe a 100 times to its current state, only once would 
that state have 
an active technological species on one of the habitable planets in the observable universe.
Again, the resulting lower limit may not rule out the existence of humanity.

Statistical inference based on low statistics counting measurements is well established. 
The parameters from Eq.~\ref{eq:drake3} are adopted.
The number of communicating civilizations $n_{\rm obs}$ that occur in the observable universe follows the Poisson distribution and depends only on $n^o_{\rm civ}$, given that $N^o_H$ is large ($\gg\! 100$) and assuming that $p_L$ is small ($< \! 0.1$).
The probability of at least one such civilization equals one minus the chance of no occurrence on any habitable planet:
\begin{eqnarray} \label{eq:nonzero}
    P(n_{\rm obs}\ge 1|\, p_L,\, N^o_{H}) &=& 1 - (1-p_L)^{N^o_H} \nonumber \\
    &=& 1 - \Big(1-\frac{n^o_{\rm civ}}{N^o_H}\Big)^{N^o_H} \nonumber \\
    &\simeq& 1 - e^{-n^o_{\rm civ}} \,,
\end{eqnarray}
where 
the estimate $n^o_{\rm civ}$ is the total expectation value.

The same holds when considering different classes of habitable planets ($i$), with $n^o_{\rm civ} = \sum_i p_{L,i}\,N^o_{H,i}\,$:
\begin{eqnarray} \label{eq:nonzero_i}
    P(n_{\rm obs}\ge 1|\, \{p_{L,i},\, N^o_{H,i}\}) &=& 1 - \prod_i\, (1-p_{L,i})^{N^o_{H,i}} \nonumber \\
    &\simeq& 1 - e^{- \sum_i p_{L,i}\,N^o_{H,i}}\,. 
\end{eqnarray}

Stated differently, $P(n_{\rm obs}\ge 1|\, n^o_{\rm civ}) = 1 - e^{-n^o_{\rm civ}}$.
For example, $P(n_{\rm obs}\ge 1|\, n^o_{\rm civ}=1) = 63\%$.
For $n^o_{\rm civ}\ll1$, to good approximation $P(n_{\rm obs}\ge 1|\, n^o_{\rm civ}) \approx n^o_{\rm civ}$.

With at least one occurrence,
from Eq.~\ref{eq:nonzero}
the lower limit on $n^o_{\rm civ}$ at 95\% C.L. is $0.051$.
Based on the ratio of stars, for the Galaxy this corresponds to 
the lower limit on $n^g_{\rm civ} > 8 \times 10^{-13}$ at 95\% C.L..

This implies that any model predictions below this limit, e.g. from the Rare-Earth hypothesis, are statistically excluded.

\section{Implications of the Lower Limit}

To demonstrate the impact of the pessimistic lower limit, part of the results of \citet{sandberg2018dissolving} have been reproduced here.
The authors
include estimated uncertainties on all factors in Eq.~\ref{eq:drake1}, without correlations, and sample the corresponding values of $n^g_{\rm civ}$.
For completeness, the prior distributions are summarized in Table~\ref{tab:uncertainties}.
The resulting distribution of $n^g_{\rm civ}$ is reproduced in Fig.~\ref{fig:sandberg} as a prior predictive check.
The prediction camps $n^g_{\rm civ}\le 1$ and $n^g_{\rm civ} \sim R^{*} L$ are both expressed in the distribution.
Most of the exponential tail towards small values is due to the modeling choice behind $f_l$, reflecting the large uncertainty on this parameter.
Together the uncertainties in Table~\ref{tab:uncertainties} cover many orders in magnitude\footnote{In view of the old-evidence argument, the priors in Table~\ref{tab:uncertainties}, together ranging from non-existence to abundant intelligent life in the observable universe, are interpreted as prior to the occurrence of evidence, and not as posterior to the development of humankind.};
the tail has been truncated in the figure, but has values smaller than $10^{-100}$ in our setup.

\begin{table}[!t]
\begin{minipage}[b]{0.38\linewidth}
  \centering
  \begin{tabular}{lll}\toprule
    \textit{Symbol} &  & \textit{Distribution} \\ \midrule
    $R^*$ &  & LogUniform($1$, $100$) \\
    $f_p$ &  & LogUniform($0.1$, $1$) \\
    $n_e$ &  & LogUniform($0.1$, $1$) \\
    $f_l$ &  & $1$ $-$ exp ($-\lambda Vt$) \\
          &  & $\lambda Vt\sim$ LogNormal($1$, $50$) \\
    $f_i$ &  & LogUniform($0.001$, $1$) \\
    $f_c$ &  & LogUniform($0.01$, $1$) \\
    $L$ &  & LogUniform($10^2$, $10^{10}$) \\ \bottomrule
  \end{tabular}
  \vspace{4mm}
  \caption{Parameters ranges used.}
  \label{tab:uncertainties}
\end{minipage}\hfill
\begin{minipage}[b]{0.6\linewidth}
\centering
  \includegraphics[width=\linewidth]{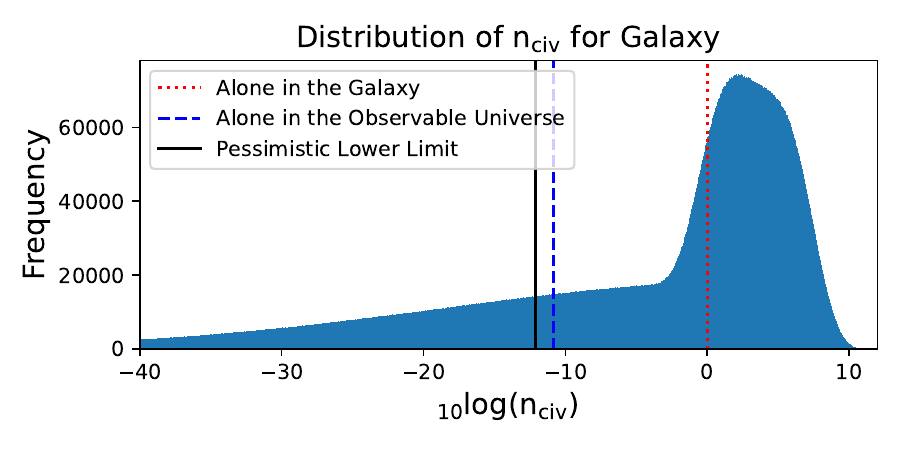}
  \captionof{figure}{Distribution of $n^g_{\rm civ}$ when including uncertainties on all factors in Eq.~\ref{eq:drake1}, reproduced from \citet{sandberg2018dissolving}.}
  \label{fig:sandberg}
\end{minipage}
\end{table}

The authors take 
the Drake equation to predict the total number of active technological civilizations, but life on Earth is implicitly assumed to be uninformative regarding abiogenesis on other worlds.
They point out that, including the uncertainties, significant fractions of the generated values of $n^g_{\rm civ}$ suggest that humanity is alone in the Galaxy and the observable universe. 
See Table~\ref{tab:alone} for the resulting fractions of the parameter space. 
These numbers are reproduced reasonably well in our study.\footnote{The tail in the distribution of $n^g_{\rm civ}$ in \citet{sandberg2018dissolving} is remarkably flat, resulting in slightly higher fractions of being alone.}

Including the obtained pessimistic lower limit, the tail below the limit is statistically excluded.
The parameter space that humanity is alone in the Galaxy remains sizable, down from 48\% to 32\% in our study.
However, for the observed universe the fraction that humanity is alone
drops substantially, from 26\% to 2.4\%.
Stated differently, 97.6\% of the parameter space has $n^o_{\rm civ} > 1$ given the evidence of life on Earth.

This changes the conclusion of the paper in a subtle but relevant way.
We may still be alone in the Galaxy, but the odds of not being alone in the observable universe are now much more favorable.

For Poisson statistics, if one simply requires that at least one event has occurred with reasonably high probability, the lower bound on $n^o_{\rm civ}$ is approximately 1. 
This is an intuitive result: if you see one event, the expected number of events is likely 1 or more.
%
%
In the absence of other detected civilizations, and treating humanity as a single data point, the most statistically straightforward estimate for the expected number of events is $n^o_{\rm civ} = 1$. 
While other priors could be argued, adopting $n^o_{\rm civ} = 1$ serves as a baseline, conservative estimate grounded in the available evidence. 
We use this value to illustrate the probabilistic consequences of our lower limit.

\begin{table}[!t]
\begin{minipage}[b]{0.46\linewidth}
  \centering
  \begin{tabular}{lll}\toprule
    \textit{Model} & $P(n_{\rm civ}<1)$ & $P(n_{\rm civ}<1)$ \\ 
    & in Galaxy & in Observable \\ 
    & & Universe \\ \midrule
    Sandberg et al. & 52\% & 38\% \\
    Our reestimate & 48\% & 26\% \\
    With lower limit & 32\% & 2.4\% \\ \bottomrule
  \end{tabular}
    \vspace{4mm}
  \caption{Key results from the distribution of $n^g_{\rm civ}$.}
  \label{tab:alone}
\end{minipage}\hfill
\begin{minipage}[b]{0.53\linewidth}
\centering
  \includegraphics[width=\linewidth]{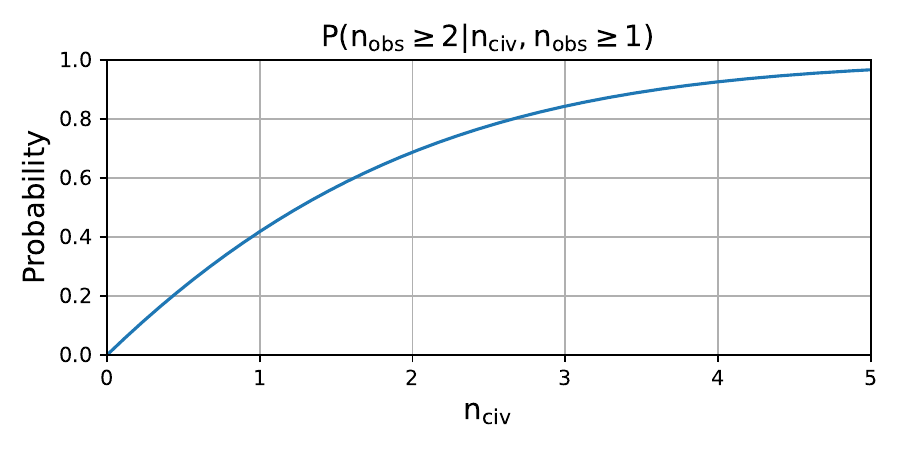}
  \captionof{figure}{Probability of existence of extraterrestrial civilizations as a function of 
$n^o_{\rm civ}$, given the existence of humankind.}
  \label{fig:poisson}    
\end{minipage}
\end{table}

Fig.~\ref{fig:poisson} shows the probability of the existence of other communicating civilizations as a function of 
$n^o_{\rm civ}$, given that at least one event has been observed:
\begin{eqnarray} \label{eq:condprod}
P(n_{\rm obs}\ge 2 | n^o_{\rm civ},\, n_{\rm obs}\ge 1) &=& \frac{P(n_{\rm obs}\ge 2 | n^o_{\rm civ})}{P(n_{\rm obs}\ge 1| n^o_{\rm civ})} \nonumber \\
&=& \frac{1 - (1+n^o_{\rm civ})\,e^{-n^o_{\rm civ}}}{1 - e^{-n^o_{\rm civ}}} \,.   
\end{eqnarray}
The formula is sensitive to values of $n^o_{\rm civ}$ around 1.
For $n^o_{\rm civ} < 1$ it approximates to $n^o_{\rm civ}/2$.
For $n^o_{\rm civ} \gg 1$ the probability quickly goes to 100\%.
The values of $n^o_{\rm civ} = \{0.051, 0.5, 1, 2, 4\}$ give probabilities of $\{2.5, 22.9, 41.8, 68.7, 92.5\}\%$ respectively.
%

The takeaway message:
the odds of having a second technological civilization given the occurrence of a first one and a necessarily compatible expectation value are relatively high.
Even with the baseline estimate of $n_{\rm civ}^o = 1$ there is a probability of 42\% for the existence of other communicating civilizations in the observable universe.


\section{Discussion and Limitations}

The results in this paper are conditional on the acceptance 
of the premise that the existence of humanity on Earth is predictively informative.
While we think the old-evidence argument is logical and sound,
we acknowledge that the validity of applying this argument to the anthropic principle is a subject of statistical and philosophical discussion.
%
%

In line with the old-evidence argument, we have argued that 
models that predict $n^g_{\rm civ}$ or $n^o_{\rm civ}$ need to be updated to include the pessimistic lower limit from the occurrence of humanity.
In particular, models with $n^o_{\rm civ}\!\ll\! 1$ are impacted, and to a lesser extend those with $n^g_{\rm civ}\!\ll\! 1$. 

The pessimistic lower limit refines our understanding of the possible existence of extraterrestrial life in a subtle way. 
At the Galaxy level, the impact of the lower limit is limited.
The Rare Earth hypothesis is not ruled out, only part of its phase space is excluded. 
As shown, we may still be alone in the Galaxy;
however, any such model may not statistically exclude the existence of humanity.
The odds of being alone in the observable universe are now disfavored,
shifting the focus to the model camp of $n^o_{\rm civ} \ge 1$.
We emphasize that even with the low-end estimate of $n^o_{\rm civ}\! =\! 1$ there is a decent chance for the existence of other technological civilizations.

When decomposing the Drake estimate into a sum of contributions from Earth-like and other types of habitable planets, each class with different parameters, it should be understood that while the lower limit still applies to the sum, it comes from the former class only.
The parameters of any other classes are  unconstrained, meaning 
their contributions to $n^o_{\rm civ}$ could still be as low as zero. 

This study has focused exclusively on establishing a lower limit on $n^o_{\rm civ}$ based on humanity's existence. 
Besides the minimalist baseline value of $n^o_{\rm civ}\! =\! 1$, 
no predictions are made for $n^g_{\rm civ}$ or $n^o_{\rm civ}$ or their possible ranges. 
%
%
The distribution of $n^g_{\rm civ}$ in Fig.~\ref{fig:sandberg} strongly depends on the modeling choices for the uncertainties on the parameters of the Drake equation.
%
%
%
We have not incorporated the constraining effect of the Fermi observation, i.e. the lack of any detected extraterrestrial signals. 
Including this observation, as done in~\cite{sandberg2018dissolving}, can constrain the upper end of the distribution of $n^g_{\rm civ}$ (or $n^o_{\rm civ}$), in a model-dependent way. 
A combined analysis can therefore narrow the probable range of $n^g_{\rm civ}$ (or $n^o_{\rm civ}$) from both below and above. 
While such a combined assessment is left for future work, the lower limit established here
and the resulting restriction of the Drake equation parameter space
remain statistically robust on its own terms.


\section{Conclusion}

Following Whitmire's argument that life on Earth is predictively relevant about the probability of abiogenesis on other Earth-like planets, 
this work treats the Drake equation as an estimate of all technological civilizations in a statistical counting experiment, 
with humanity as informative data point, 
and explores its statistical consequences.
This sets a pessimistic lower limit of $n^o_{\rm civ} > 0.051$ at 95\% C.L. for the observable universe, or $n^g_{\rm civ} > 8\times10^{-13}$ at 95\% C.L. for the Galaxy.

Including this limit, 
the Rare-Earth hypothesis for the Galaxy is not ruled out, although the lower end of the phase-space is constrained. 
However, 
the odds of being alone in the observable universe are now disfavored, 
our study finds $P(n^o_{\rm civ}>1 | n_{\rm obs} \ge 1) = 97.6\%$.
Using the low-end estimate of $n^o_{\rm civ} = 1$ gives a 42\% probability for the existence of other communicating civilizations in the observable universe.

We find these results to be 
good arguments for continued searches for extraterrestrial life.



\section*{Acknowledgements}

We thank Ilan Fridman for carefully reading the manuscript, and Dave and Stefan, our hairdressers, for insightful discussions on the existence of extraterrestrial life. 

\bibliographystyle{unsrtnat}
\bibliography{references}  






\end{document}